February 24, 2012

# The Transition between Stochastic and Deterministic Behavior in an Excitable Gene Circuit

Robert C. Hilborn*, Benjamin Brookshire, Jenna Mattingly, Anusha Purushotham, and Anuraag Sharma

The University of Texas at Dallas
Richardson, TX 75080, USA

**Abstract**

We explore the connection between a stochastic simulation model and an ordinary differential equations (ODEs) model of the dynamics of an excitable gene circuit that exhibits noise-induced oscillations. Near a bifurcation point in the ODE model, the stochastic simulation model yields behavior dramatically different from that predicted by the ODE model. We analyze how that behavior depends on the gene copy number and find very slow convergence to the large number limit near the bifurcation point. The implications for understanding the dynamics of gene circuits and other birth-death dynamical systems with small numbers of constituents are discussed.

## Introduction

Gene circuits are sets of interacting genes and proteins (and perhaps other biological molecules). It is now widely recognized that stochastic fluctuations play an important role in the dynamics of gene circuits [1]. The effects of these fluctuations on gene expression have been studied in a variety of papers [2-7]. In fact, these stochastic fluctuations may explain some aspects of phenotype behavior: how differentiated cells emerge from cells with identical genetic makeup and identical environments, although many other so-called epigenetic effects such as DNA methylation, histone modification, and small interfering RNAs also play a role in differentiation and inheritance of differentiated characteristics [8,8,9].

These fluctuations, always present when gene copy numbers and the number of resulting messenger RNAs (mRNAs) and proteins are small, must be taken into account to understand the dynamics of genetic oscillators such as circadian clock networks. Similar issues arise in the modeling of chemical reaction networks [10] and ecological populations [11] when the number of constituents is small. In this paper, however, we focus on the dynamics of gene circuits.

Many studies of gene regulatory circuits have focused only on steady-state behavior. For many gene circuits, however, temporal behavior yields important information that is not accessible from just steady-state conditions. Furthermore, in many situations, protein production occurs in bursts and sometimes gene regulation varies in time due to environmental changes, cell differentiation and disease. Measuring and understanding the temporal dynamics of gene circuits also helps to identify causal relations and feedback loops, the details of which are hidden under steady-state conditions. The importance of temporal behavior in understanding gene circuits was emphasized in a recent review [12].

Many important cellular and organismal periodic processes are controlled by genetic networks with more or less periodic oscillations. From the dynamics point of view, these periodic oscillations are surprising because most genes are present with only small copy numbers—typically one or two copies per cell. Naively, one might expect that the large relative fluctuations normally associated with small molecular numbers would lead to irregular oscillations. In reality, many of these genetic circuits exhibit quite regular periodicity even when the copy numbers are small. The long-term goal of our study is to understand how genetic circuits are able to maintain regular oscillations in spite of the molecular fluctuations associated with small numbers. In this paper, we focus on the connections between two classes of models of gene circuit dynamics: deterministic (ordinary differential equation) models and stochastic models.

Once the key elements of a genetic circuit are identified, the dynamics of the circuit can be specified either by a set of deterministic ordinary differential equations (ODEs) (often called rate equations) or by a stochastic formulation, usually implemented as a Monte Carlo simulation of the dynamics. The stochastic formulation, by design, includes fluctuations due to small molecular numbers. Such fluctuations are of course absent from the ODE models. The goal of this paper is to see how the behavior of stochastic models is related to the deterministic behavior of the commonly used ODE models.

An equivalent stochastic formulation using the so-called master equation [13] for the dynamics of the probability distribution for the number of molecules of the relevant species is, in most cases, intractable for anything but the simplest networks. There are also intermediate methods that add stochastic terms to the deterministic differential equation models. These intermediate methods are often described by chemical Langevin equations [14], which can be derived from the master equation for the probability distribution under appropriate approximations. While the stochastic models are viewed as being more realistic than the rate equation models for gene circuits, they are computationally expensive even for modest size networks. Thus, we would like to understand when the ODE results can be used in place of the stochastic models and when they cannot.

The usual mathematical folklore is that the behavior described by the stochastic simulations should approach that described by the rate equations as the number of molecules, including the number of gene copies, becomes large. What is generally lacking is any prediction of how large those numbers must be to see similar behavior. In the absence of analytical solutions for the probability distributions that result from the chemical master equation, we must resort to simulations and phenomenological models, which we shall pursue in this paper.

The converse problem, how the deterministic results are modified as the number of genes and molecules gets smaller, can be treated in a systematic fashion via the so-called omega-expansion of the master equation as developed by van Kampen [13]. This approach yields, as a first approximation, the rate equations for the system dynamics. The next terms in the expansion (proportional to $1/\sqrt{N}$ , where $N$ is a measure of the system size) give a Langevin equation description of the dynamics: rate equations supplemented with stochastic driving terms.

For realistic biological systems, however, we are almost always concerned with small gene copy numbers and relatively small numbers of mRNAs and proteins. The question then arises of how the dynamics are modified as these numbers increase though they remain far from the traditional "thermodynamic (large number) limit."

The gene circuit model used here falls into the class of "excitable" dynamical systems: for a range of parameter values, the behavior of the system tends toward a time-independent steady state (after initial transients die away). However, a sufficiently large perturbation can push the system away from its steady state conditions and excite a large excursion (a protein production burst, for a gene circuit) before the system returns to the steady state. For an excitable gene circuit, we have found as expected (the details are given in what follows) that as the gene copy number increases, the molecular (stochastic) model results, in general, approach those of the rate equation (ODE) model as long as the parameter values are significantly different from those on the border between oscillations and steady-state behavior in the rate equation model. Near the boundary between the two dynamical regimes, however, the stochastic model behavior is oscillatory with a regularity that is almost independent of gene copy number (at least over the range of gene copy numbers [between 1 and 48] explored in this study). This is a new result that may have implications for the evolutionary interpretation of the design of genetic oscillators: there may be an advantage in having oscillating genetic networks poised on the boundary (as defined by rate equations) between oscillatory and steady-state behavior because in that regime, the oscillator properties are reasonably independent of gene copy number. This behavior might be studied experimentally using synthetic gene networks [15] [16].

What remains missing is a method for predicting how the convergence to the deterministic behavior depends on the numbers of the various molecular constituents and the parameter values. For an excitable system, the answer to this question should depend on the parameters of the dynamical model that determine the height of the escape barrier that sets the excitability of the system's dynamics. Later in this paper, we provide a phenomenological model that may guide more formal analytical treatments.

The effects of gene copy number variation are of interest more broadly because it has been recognized that such variations lead to phenotypic variation and, in some cases, disease. In a recent paper [17], the effects of deleting one copy of various genes in the galactose response system in the yeast *Saccharomyces cerevisiae* were studied and modeled to explore if and how the system compensates for changes in "network dosage" (essentially changes in gene copy number). Copy number variations in the human genome have been explored in recent studies [18] [19].

An extreme case occurs in aneuploidy [20], which refers to having an abnormal number of chromosomes. Aneuploidy can affect health and disease as well as phenotypic variations [21]. Experiments have shown that cells do not generally compensate for changes in gene copy or chromosome number [22]. Gene copy number variations may also result in nervous system disorders [23].

In recent work [24], Zakharova et al have investigated system-size effects in cellular network models of oscillatory gene circuits. The models allow coupling among the gene circuits in different cells. Their work focused on seeing how the so-called stochastic bifurcations [25] change as the number of cells in the network changes. Stochastic bifurcations are marked by changes in the structure of the probability distributions for the proteins produced by the gene circuits. For the model used, the authors found that the stochastic bifurcations were similar for networks with 1, 2, and 500 cells.

## Methods

### Oscillatory Gene Circuit Model

Since most real gene circuits are complex with many interacting genes and proteins, we have focused on a relatively simple model of a genetic oscillator [26], which we shall refer to as the VKBL model. This model involves two genes, their promoter regions, the messenger RNAs, and two product proteins, one of which enhances the production of both proteins while the other forms a complex with the activator protein thus effectively inhibiting the production. In an earlier study, Hilborn and Erwin (2008) found that the regularity of the oscillations in this model shows a local maximum as a function of gene copy number, an effect known as stochastic coherence (or coherence resonance). This was the first systematic study of stochastic coherence in a gene circuit model. In the current study, we have explored a wider range of parameter values to understand the conditions under which stochastic coherence occurs in this model and to explore the connections between the molecular (stochastic chemical reaction) model of the dynamics and the rate equation (ODE) model of the dynamics. In the stochastic model version, we have found oscillations over a much wider range of parameters, including those for which the differential equation model predicts only a steady-state. (This fact was pointed by Vilar et al. but not explored in detail.) Our results show that the rate equation predictions must be interpreted carefully and that the presumably more realistic stochastic model often shows dramatically different behavior.

The VKBL model is described in terms of 16 reactions:

$$\mathrm{D_A} + \mathrm{A} \xrightarrow{\gamma_A} \mathrm{D'_A} \tag{1}$$

$$\mathrm{D'_A} \xrightarrow{\theta_A} \mathrm{D_A} + \mathrm{A} \tag{2}$$

$$\mathrm{D_R} + \mathrm{A} \xrightarrow{\gamma_R} \mathrm{D'_R} \tag{3}$$

$$\mathrm{D'_R} \xrightarrow{\theta_R} \mathrm{D_R} + \mathrm{A} \tag{4}$$

$$\mathrm{D'_A} \xrightarrow{\alpha'_A} \mathrm{D'_A} + \mathrm{M_A} \tag{5}$$

$$\mathrm{D_A} \xrightarrow{\alpha_A} \mathrm{D_A} + \mathrm{M_A} \tag{6}$$

$$\mathrm{M_A} \xrightarrow{\delta_{M_A}} \mathrm{X} \tag{7}$$

$$\mathrm{M_A} \xrightarrow{\beta_A} \mathrm{M_A} + \mathrm{A} \tag{8}$$

$$\mathrm{A{+}R} \xrightarrow{\gamma_C} \mathrm{C} \tag{9}$$

$$\mathrm{A} \xrightarrow{\delta_A} \mathrm{Y} \tag{10}$$

$$\mathrm{D'_R} \xrightarrow{\alpha'_R} \mathrm{M_R} + \mathrm{D'_R} \tag{11}$$

$$\mathrm{D_R} \xrightarrow{\alpha_R} \mathrm{M_R} + \mathrm{D_R} \tag{12}$$

$$\mathrm{M_R} \xrightarrow{\delta_{M_R}} \mathrm{Z} \tag{13}$$

$$\mathrm{M_R} \xrightarrow{\beta_R} \mathrm{M_R} + \mathrm{R} \tag{14}$$

$$\mathrm{C} \xrightarrow{\delta_A} \mathrm{R} \tag{15}$$

$$R \xrightarrow{\delta_R} W \qquad (16)$$

In Eqs. (1)-(16), the non-italic symbols represent the specific molecule type (rather than the number of molecules). $D'_A$ and $D_A$ represent the DNA operator sites with and without protein A (Activator) bound, respectively. $D'_R$ and $D_R$ are the corresponding R (Repressor) protein operator sites. $M_A$ and $M_R$ are the mRNAs for the two proteins. C represents the Activator-Repressor complex. The model assumes that when the complex decomposes Activator is degraded. Thus, $\delta_A$ appears in Eq. (15). W, X, Y, and Z are inactive decay products. The model ignores any changes in concentration due to cell growth or cell division (mitosis). The numerical values of the parameters are given in Appendix A Table 1; they are the same as those used by [26]. A schematic diagram of the VKBL model is shown in Fig. 1.

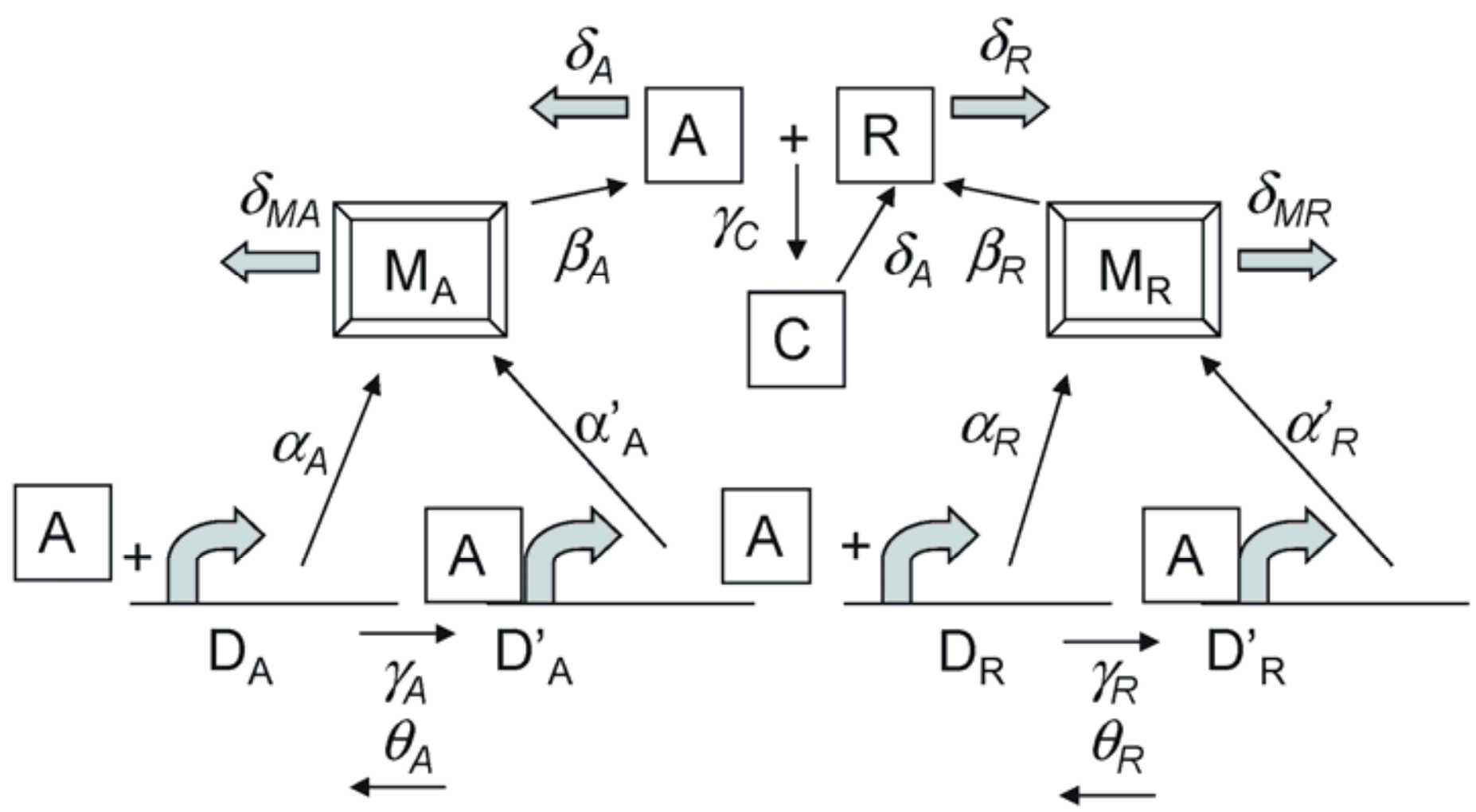

**Figure 1. A schematic diagram of the VKBL model**. The symbols next to the reaction arrows are the reaction rate parameters. The other symbols are defined in the text. Modified from a similar figure in Vilar et al 2002.

The dynamics predicted by the 16 reactions can be simulated by means of the Gillespie algorithm [10,27], which makes use of computer-generated random numbers to implement the reactions stochastically in such a way that times between reactions of a particular type follow an exponential distribution and the probability of a reaction's occurring is proportional to its reaction rate constant. Figure 2 shows the dynamics of the R protein for the R protein degradation rate $\delta_R = 0.2$ .

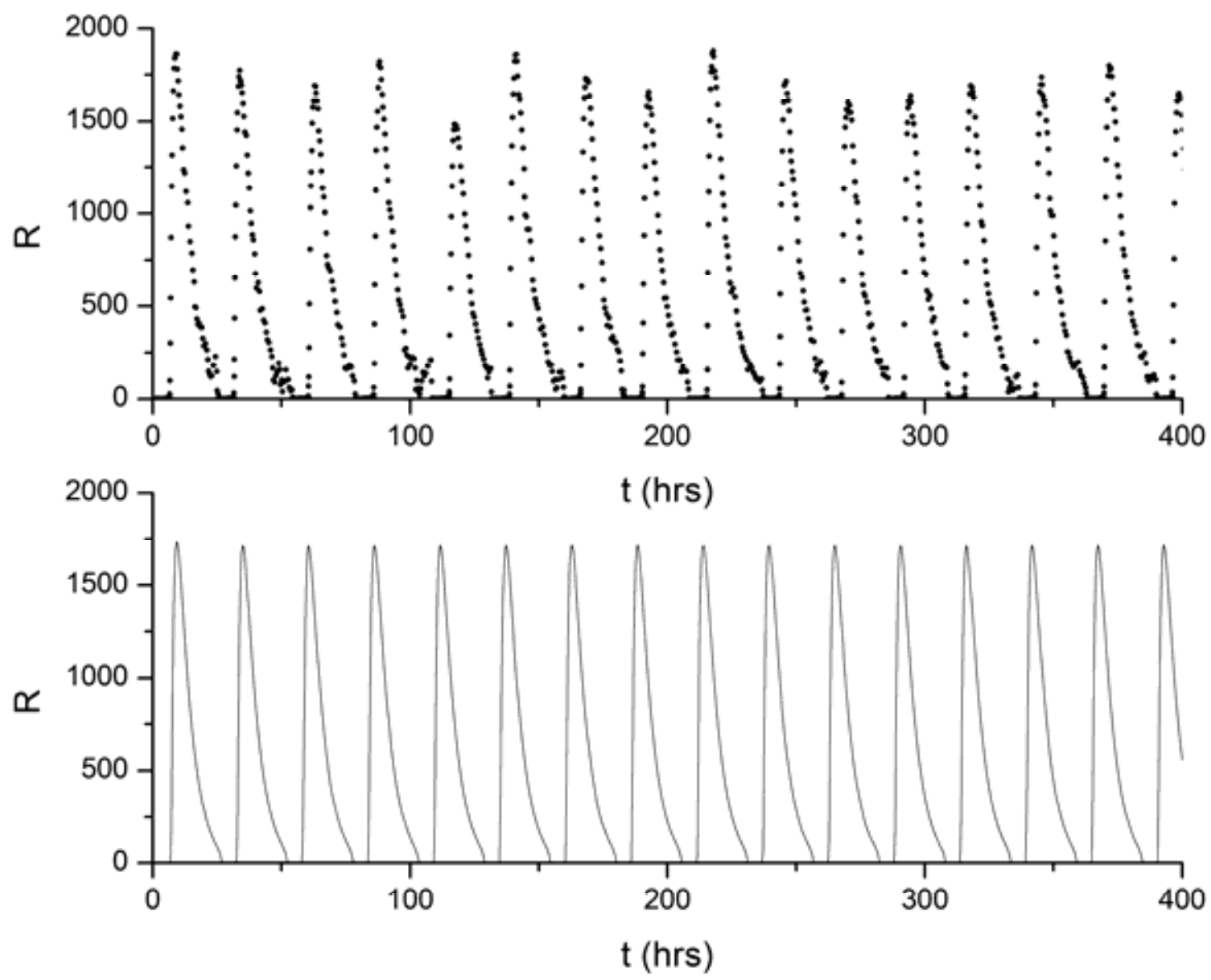

**Figure 2. R protein number as a function of time (in hours).** Upper panel: R protein degradation rate $\delta_R = 0.2$ (and other parameters as described in the text) with the dynamics simulated using the stochastic algorithm (Gillespie model). Gene copy = 1. The corresponding rate equation model predicts perfectly periodic oscillations, as shown in the lower panel, for this set of parameters.

The VKBL model can also be described by nine rate equations for the number (or equivalently, concentrations, since we are dealing with a fixed volume) of bound and unbound operator sites, mRNAs, and the resulting proteins and the protein complex. Following the notation of [26], we write the rate equations as

$$dD_A / dt = \theta_A D'_A - \gamma_A D_A A \tag{17}$$

$$dD_R / dt = \theta_R D'_R - \gamma_R D_R A \tag{18}$$

$$dD'_A / dt = \gamma_A D_A A - \theta_A D'_A \tag{19}$$

$$dD'_R / dt = \gamma_R D_R A - \theta_R D'_R \tag{20}$$

$$dM_A / dt = \alpha'_A D'_A + \alpha_A D_A - \delta_{M_A} M_A \tag{21}$$

$$dM_R / dt = \alpha'_R D'_R + \alpha_R D_R - \delta_{M_R} M_R \tag{22}$$

$$\begin{aligned} dA / dt = \beta_A M_A + \theta_A D'_A + \theta_R D'_R \\ - A\left(\gamma_A D_A + \gamma_R D_R + \gamma_C R + \delta_A\right) \end{aligned} \tag{23}$$

$$dR / dt = \beta_R M_R - \gamma_C AR + \delta_A C - \delta_R R \tag{24}$$

$$dC / dt = \gamma_C AR - \delta_A C, \tag{25}$$

where the italic symbols indicate the number (concentration) of molecules present of each type. The rate equation model treats the molecular numbers (concentrations) as continuous variables.

In exploring the dynamics of the VKBL model, we used $\delta_R$, the degradation rate of protein R, as the control parameter. (There is some empirical evidence that protein degradation plays a role in controlling circadian oscillation periods [28].) For $\delta_R < 0.087812\ldots$, the rate equations predict that the system will move towards a steady state with unchanging numbers of proteins A and R. For $\delta_R > 0.087812\ldots$, the behavior consists of perfectly periodic oscillations consisting of bursts of mRNA and protein production. In the language of nonlinear dynamics, $\delta_R = 0.087812...$ (with the values of the other parameters listed in Table 1) marks a subcritical Hopf bifurcation in the dynamical behavior of the ODE model. Figure 3 illustrates the bifurcation by plotting the maximum and minimum values of $R(t)$ as a function of $\delta_R$. There is no indication of hysteresis (bistability) near this bifurcation.

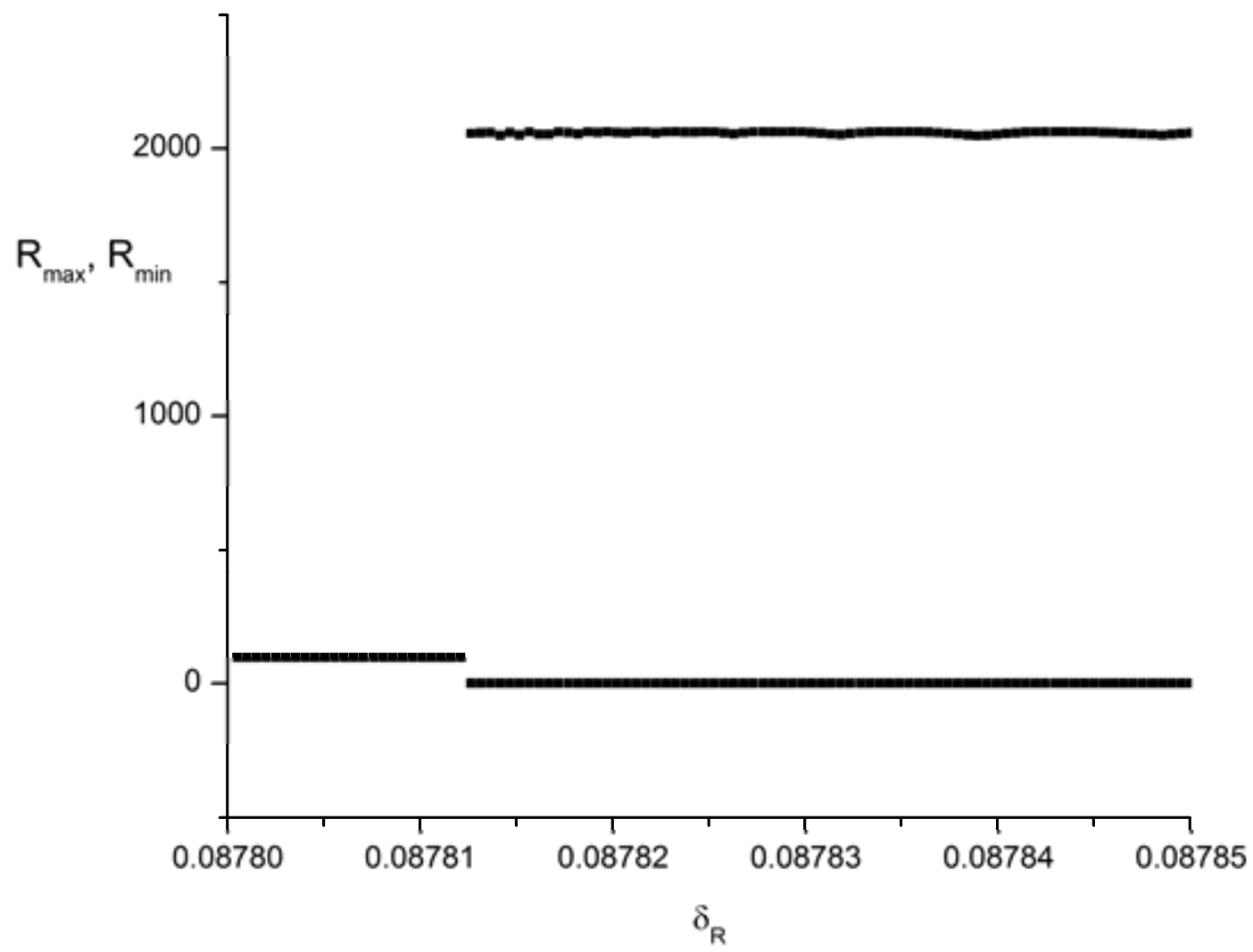

**Figure 3. A bifurcation diagram for the genetic circuit model.** The maximum and minimum numbers of protein R are plotted as a function of the R protein degradation rate $\delta_R$ with gene copy number = 1. Below $\delta_R = 0.087812...$ the system displays steady-state (time independent) behavior.

The VKBL model exhibits excitable behavior for $\delta_R < 0.087812$ : the rate equations predict that concentrations will tend to steady state values, but a sufficiently large perturbation will induce a burst of mRNA and protein production before the behavior returns to the steady state. In that range of parameter values, however, the stochastic simulations show more or less regular bursts of protein production—a type of noise-induced oscillation. The dynamics of the system can be understood by plotting the nullclines for a reduced version of the VKBL model [26]. The reduced model is a two-variable model derived by assuming that the molecular numbers of all the species except proteins R and C equilibrate to the instantaneous values of the numbers of R and C. The system is then described by two differential equations:

$$dR/dt = \frac{\beta_R(\alpha_R\theta_R + \alpha'_R\gamma_R a[R(t)])}{\delta_{M_R}(\theta_R + \gamma_R a[R(t)])} - \gamma_C a[R(t)]R(t) + \delta_A C(t) - \delta_R R(t) \quad (26)$$

$$dC/dt = \gamma_C a[R(t)]R(t) - \delta_A C(t), \quad (27)$$

where,

$$a[R(t)] = \frac{1}{2}\left(\alpha'_A \rho(R) - k_D\right) + \sqrt{\left(\alpha'_A \rho(R) - k_D\right)^2 + 4\alpha_A \rho(R) k_D} \tag{28}$$

and

$$\rho(R) = \frac{\beta_A}{\delta_{M_A}\left(\gamma_C R + \delta_A\right)} \text{ and } k_D = \frac{\theta_A}{\gamma_A} \quad . \tag{29}$$

For the reduced model, the intersection of the nullclines (the curves for which d*R*/dt and d*C*/d*t* = 0) determine the fixed point for the system. Figure 4 shows the nullclines with the *RC* plane projection of a trajectory from the full stochastic model superposed for $\delta_R = 0.06$, for which value the fixed point is stable. The ODE version of the model shows just steady state behavior (after transients die away while the stochastic model shows noise-induced oscillations.

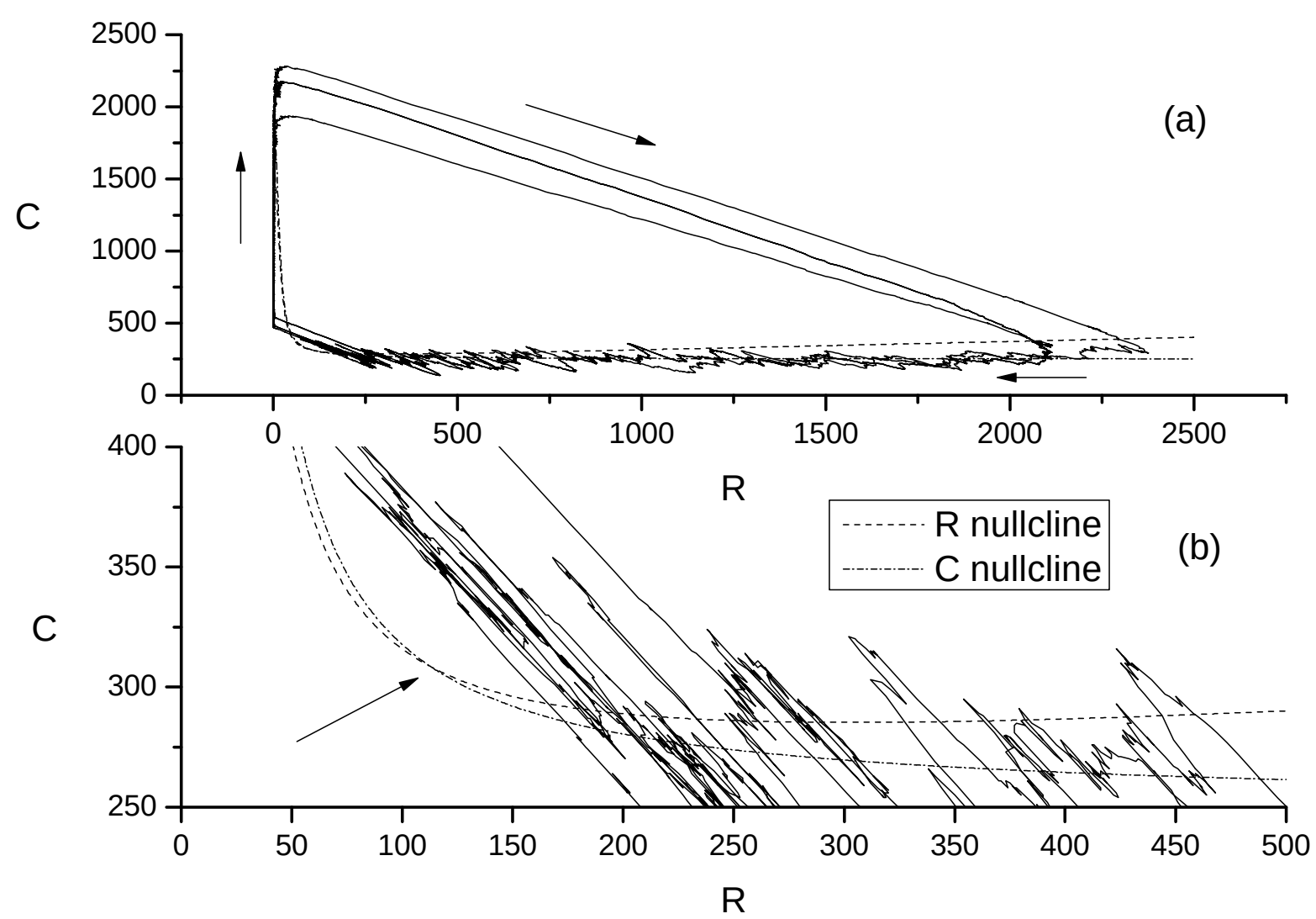

**Figure 4. A phase-space diagram for the genetic circuit model.** Panel (a): The nullclines for the reduced model described by Eqns. (26) and (27) with the R protein degradation rate $\delta_R = 0.06$ are shown by the dash and dash-dot lines. The RC plane projection of a trajectory from the full stochastic model is superposed. This segment of the trajectory consists of three protein bursts. The phase space point circulates in the direction indicated by the arrows. Panel (b): An expanded view of panel (a) in the neighborhood of the fixed point, indicated by the arrow near the lower left corner.

As an aside, we note that excitable systems may also display (relatively) small amplitude oscillations around the steady-state fixed point. See, for example, [29] and [30]. We ignore those small-amplitude oscillations here.

To investigate the connection between the behavior of the ODE model and the stochastic simulation model, we study how the average time between protein production bursts—the inter-burst interval (IBI)—and the regularity of the inter-burst intervals depends on the system size. We change the system size in two ways: (1) by varying the gene copy number and (2) by varying the transcription rates (with gene copy number fixed). We note that system-size effects in a model close to the one described here were studied by [31], in which the overall volume of the system was increased, keeping concentrations fixed. Using the chemical Langevin equation method, Hou and Zin [32] studied the volume dependence of noise effects in a simplified circadian clock network consisting of only mRNA and two proteins. The latter paper did not involve the gene copy number explicitly.

## Results

### Gene Copy Number Variation

We first examine how the system behavior depends on the gene copy number. As mentioned previously, for $\delta_R > 0.087812...$ the rate equation model predicts that protein bursts will occur at regular intervals. The stochastic simulation model exhibits similar behavior for all values of the gene copy number studied here (1 through 48). This behavior is illustrated in Fig. 5.

For $\delta_R$ values below the bifurcation value, the ODE model predicts a time-independent steady state (no protein bursts). The stochastic simulation model yields protein bursts throughout this parameter range (i.e. the stochastic model predicts noise-induced oscillations) as illustrated in Fig. 6, which shows how the time between protein bursts (the inter-burst interval, IBI) depends on gene copy number for several values of $\delta_R$, the R protein degradation rate, near the ODE bifurcation value.

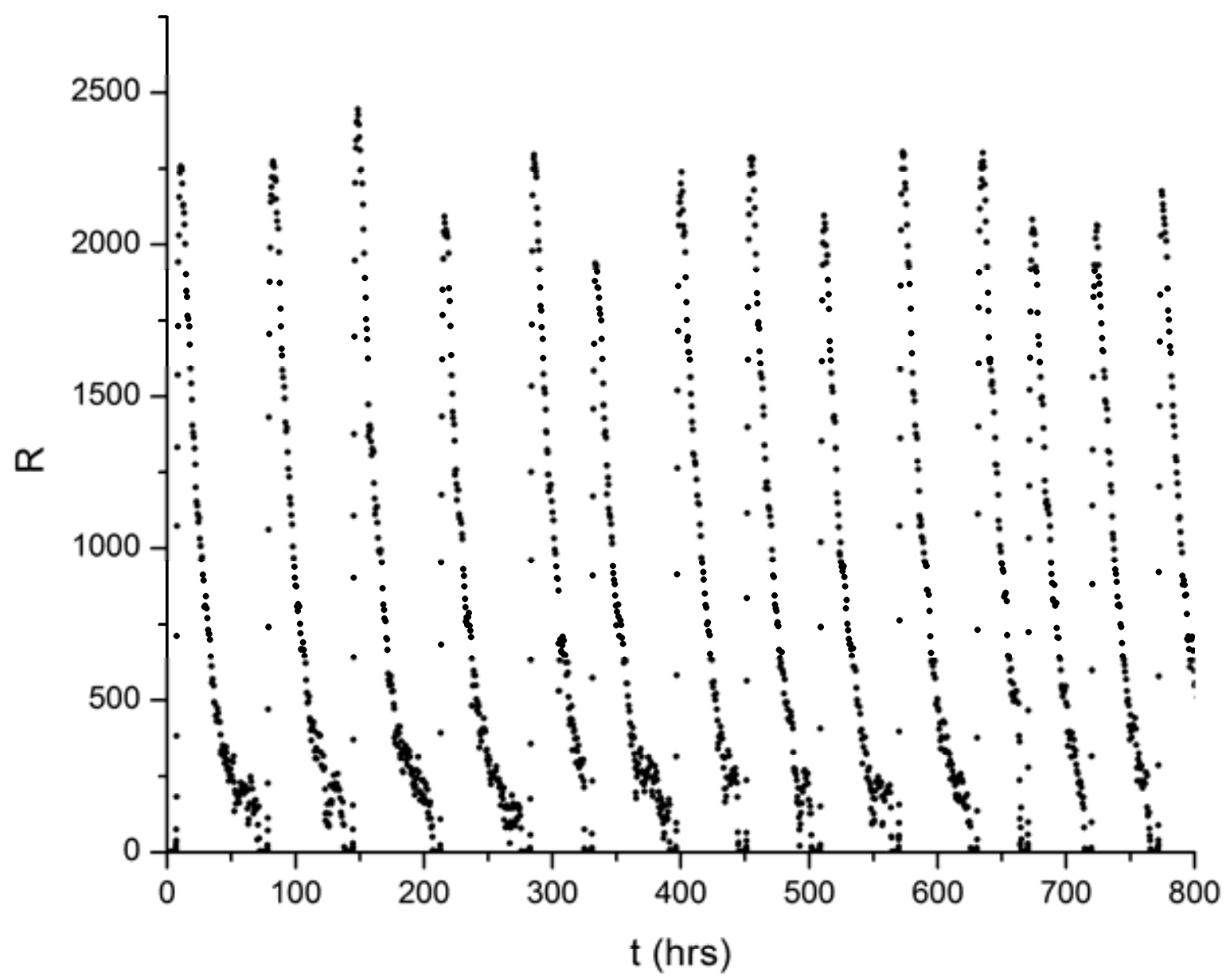

**Figure 5. R protein number as as a function of time (in hours).** The R protein degradation rate $\delta_R = 0.06$ with the dynamics calculated with the stochastic algorithm. The rate equation model predicts a time-independent steady-state (after transients die away) for this set of parameter values. Gene copy number = 1.

For $\delta_R$ values well above the bifurcation value, the IBIs are almost independent of gene copy number. This behavior is expected because the ODE model gives perfectly periodic oscillations for those parameter values and we anticipate that the stochastic simulation results should approach those of the ODE model as the gene copy number increases. There is an overall trend as well: as $\delta_R$ increases, the time between bursts decreases since the protein degradation rate is the dominant effect in setting the length of the protein burst.

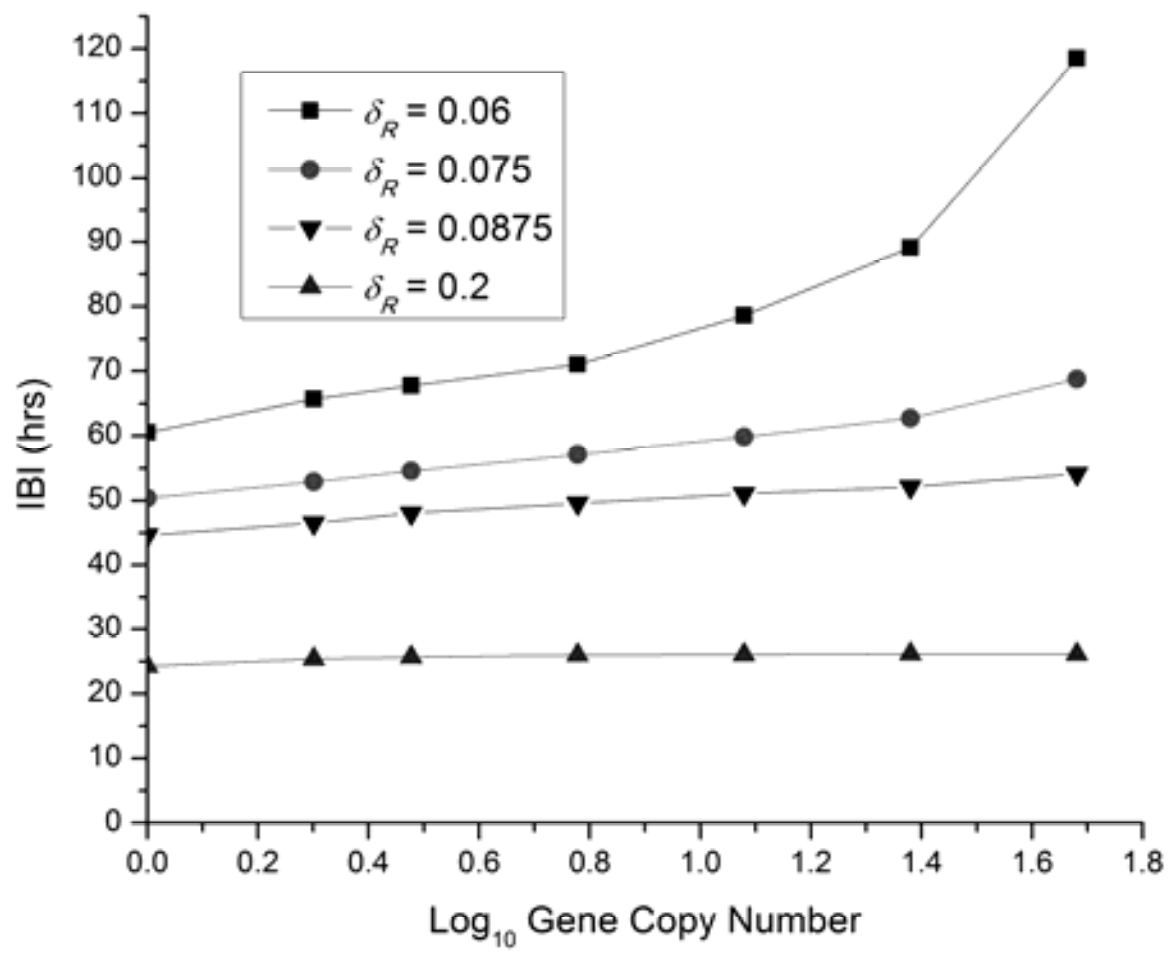

**Figure 6**. **The inter-burst interval (IBI) plotted as a function of the logarithm of the gene copy number**. Four values of the control parameter $\delta_R$, the R protein degradation rate, were used with the stochastic simulation model. Lines have been added to guide the eye.

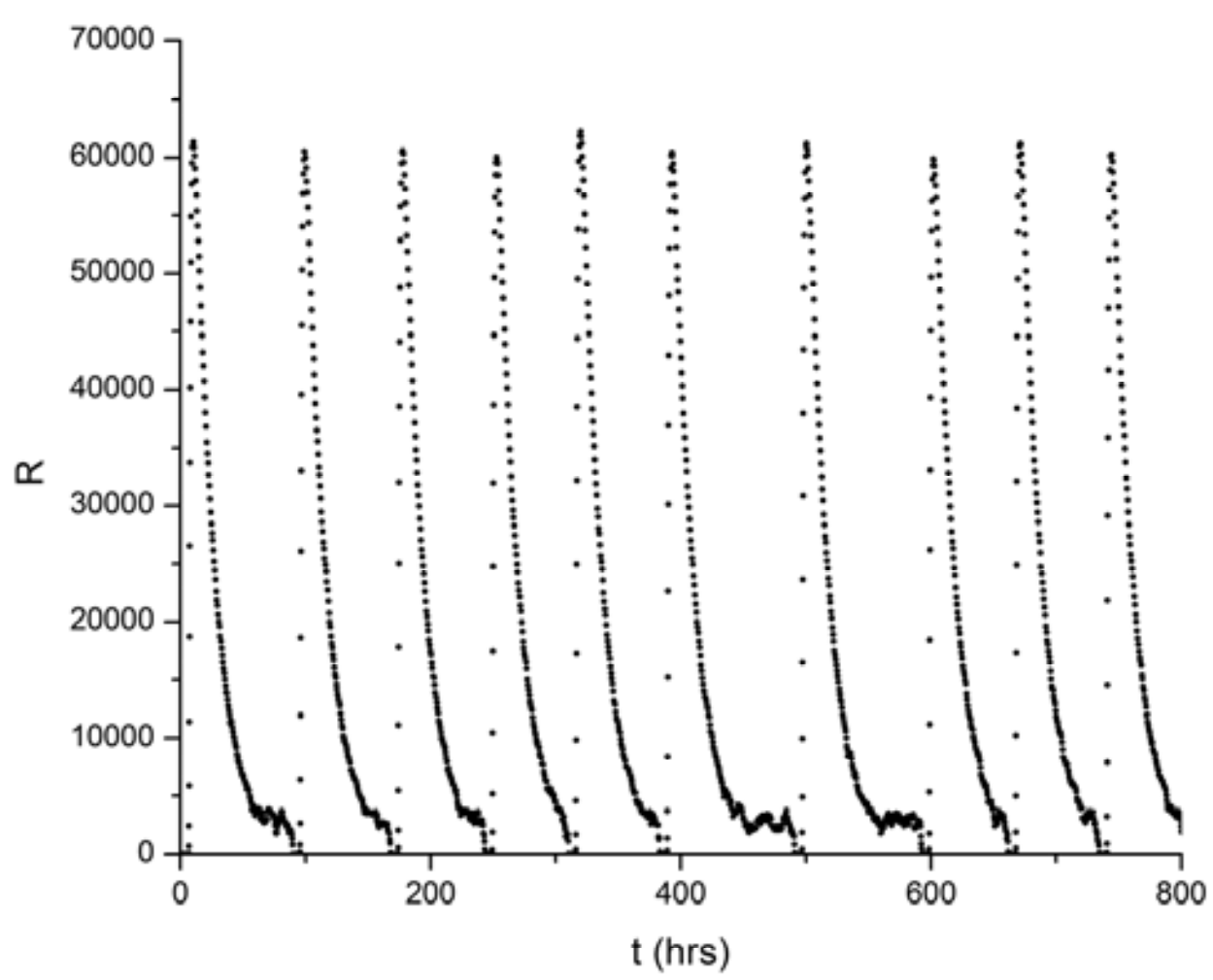

**Figure 7**. **Protein R number versus time (in hours).** Calculated using the stochastic simulation model. The R protein degradation rate $\delta_R = 0.06$ with gene copy number = 24. Compare with Fig. 5 where gene copy = 1.

For parameter values close to and below the ODE bifurcation value $\delta_R = 0.087812...$, the IBI increases as the gene copy number increases as seen in Fig. 6. The behavior begins to approach that of the ODE model (steady state) with (on average) longer times between bursts. In addition, the time between bursts becomes more irregular, as discussed below, and illustrated in Fig. 7. The amplitude of the bursts, however, becomes more regular (in terms of relative fluctuations) compared to the behavior with gene copy number = 1 (Fig. 5).

We characterize the regularity of the protein bursts with a regularity parameter $\mathcal{R}$ defined as the ratio of the average IBI ($T$) to the standard deviation of the IBIs:

$$\mathcal{R} = \frac{\langle T \rangle}{\sqrt{\operatorname{var} T}}, \qquad (30)$$

where var $T$ is the variance of the IBIs. Figure 8 shows the regularity of the time intervals between protein bursts as a function of gene copy number for $\delta_R$ values above, near, and below the bifurcation value. We see that for $\delta_R$ well below the bifurcation value, the regularity decreases as the gene copy number increases, reflecting the increasing irregularity of the time between protein bursts.

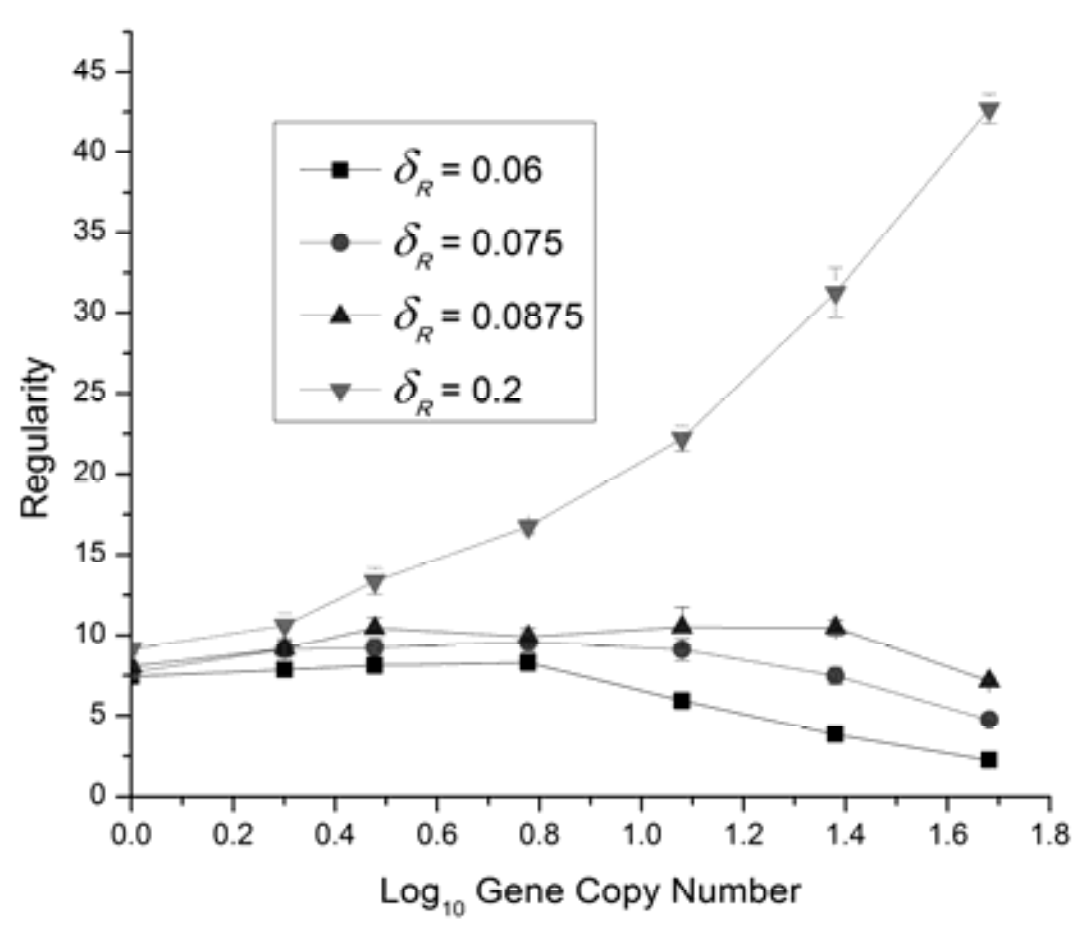

**Figure 8. The average regularity of the protein burst intervals plotted as a function of the logarithm of the gene copy number.** Four values of the control parameter $\delta_R$, the R protein degradation rate, were used with the stochastic simulation. The average was taken over 10 independent noise realizations. In most cases the uncertainty bars are smaller than the plotted symbols. Lines added to guide the eye.

For $\delta_R$ well above the bifurcation value, we see that the regularity increases rapidly with gene copy number as expected since the stochastic simulation results should approach the completely periodic ($\mathcal{R} \to \infty$) results from the ODE model for large gene copy numbers. Close to the bifurcation value, however, the regularity is almost independent of gene copy number, at least over the range of values explored here. This effect is analogous to "critical slowing down" (very long relaxation times) near phase transitions in thermodynamic systems [33,34].

## Changing transcription rates

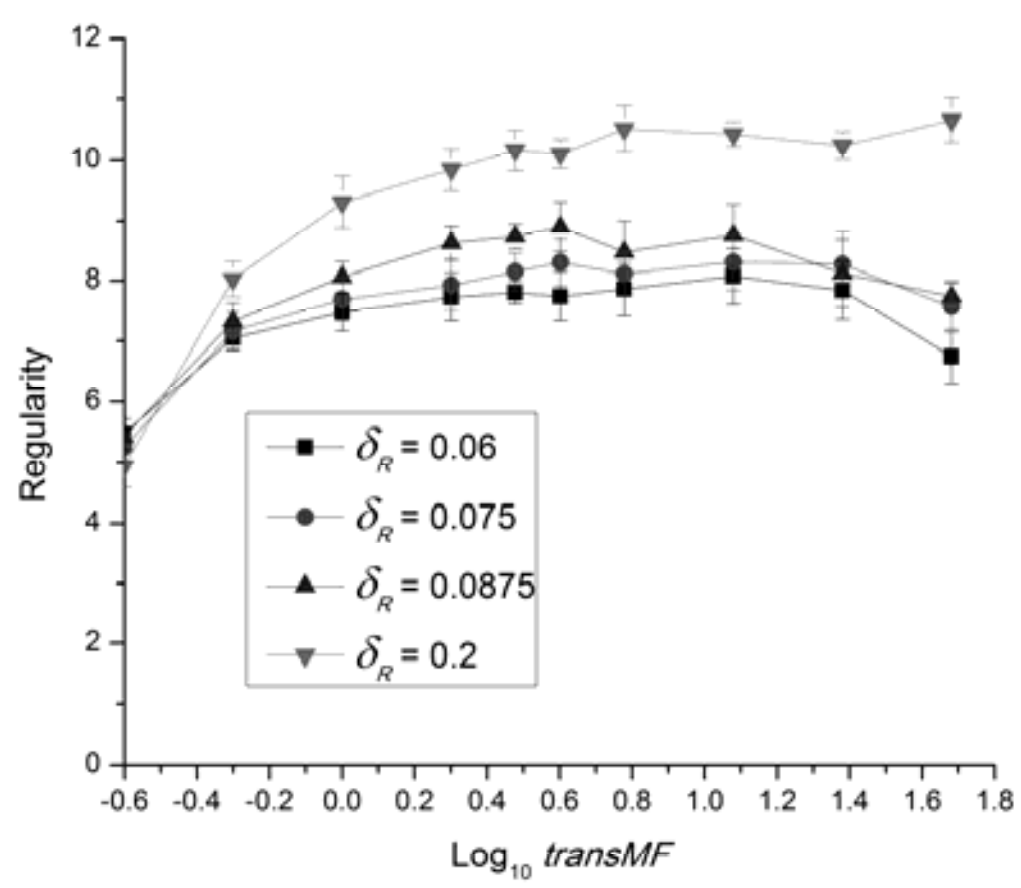

**Figure 9. The average regularity of the protein bursts as a function of the logarithm of the transcription rate multiplicative factor *transMF*.** Four different values of $\delta_R$, the R protein degradation rate, were used with the stochastic simulations. Gene copy number = 1. The averages were computed from 10 independent noise realizations.

We can also change the number of molecules in the system by increasing the transcription rates, leaving the gene copy numbers fixed. For the sake of simplicity, we implemented these increases by multiplying all of the transcription rates by a common factor *transMF*. Figure 9 displays the regularity of the protein burst intervals as a function of *transMF* for four values of the control parameter $\delta_R$, the R protein degradation rate. We see the same general trends that are shown in Fig. 8, but the rise in regularity for $\delta_R = 0.2$ is not as dramatic as that shown in Fig. 8.

We can also compare the two methods of varying system size by looking at the regularity results in cases where the product of the gene copy number and the transcription rate multiplicative factor is constant. Three such comparisons are shown in Appendix A Table 2. Note that in all three cases, the product with the higher gene copy number yields a regularity larger than that in the case with the higher *transMF*, though the differences are not large compared to the standard deviation of each of the regularity results. These results tell us that increasing the gene copy number (at least for small gene copy numbers) is more effective in enhancing the regularity than is increasing the transcription rates.

## Modeling the Interburst Interval

The results presented above came from stochastic simulations of the dynamics of the VKBL model. One might ask if a more analytic approach is available to predict the average interburst interval and the regularity of the resulting bursts. For relatively simple models of excitable systems, the answer is yes, but even then the formalism is moderately complex and may require numerical evaluation of integrals to obtain quantitative results [35].

To develop a simpler formal description of these effects, we model the stochastic dynamics as a so-called first-passage problem [13]. Let us assume that the system has settled into the state space region around the ODE fixed point. With small values of the noise amplitude, the system will remain near that fixed point. However, if a sufficiently large noise "kick" occurs, the system can escape from the neighborhood of the fixed point and undergo a large excursion through state space producing a burst of proteins before returning to the neighborhood of the fixed point (low protein numbers in the model considered here). This behavior indicates that there is an escape barrier near the fixed point, the "height" of which depends on the system parameters. A typical average first passage time $\tau$ for escape can be expressed as

$$\tau = \tau_0 \left( e^{B/STD} - 1 \right), \tag{31}$$

where $B$ is the barrier height (expressed in terms of molecule numbers), $STD$ is the noise standard deviation and $\tau_0$ is a parameter that sets the time scale for the system under study. For low and moderate noise amplitudes, the duration of the burst (the long excursion through state space) $T_B$ is approximately independent of the noise amplitude. The overall time between bursts is then given by

$$IBI = T_{\mathbf{B}} + \tau. \tag{32}$$

When $STD$ is very small compared to $B$, the average time between bursts will be large. The system hardly ever escapes from the fixed point region. As the noise amplitude increases, $\tau$ decreases: larger noise kicks allow the system to escape from the fixed point region more quickly.

The barrier height is determined by the model parameters (e.g. $\delta_R$ in the VKBL model) including the gene copy number. For typical chemical reaction models (exponentially distributed times between reactions), the noise standard deviation $STD$ is proportional to the square root of the number of molecules present (Poisson distribution). To reproduce the observed IBI as a function of gene copy number, we must have the barrier height proportional to the gene copy number raised to some power greater than 1/2. For this model, let's assume the barrier height is proportional to the gene copy number $N$ and that $STD = \sqrt{N}$, the Poisson distribution result. Eq. (32) then becomes

$$IBI = T_0 + \tau_0 (e^{a\sqrt{N}} - 1). \tag{33}$$

Figure 10 shows the results of fitting Eq. (33) to the protein interburst intervals (IBIs) calculated from the simulation data. The model provides a reasonable fit to the data over a large range of gene copy numbers.

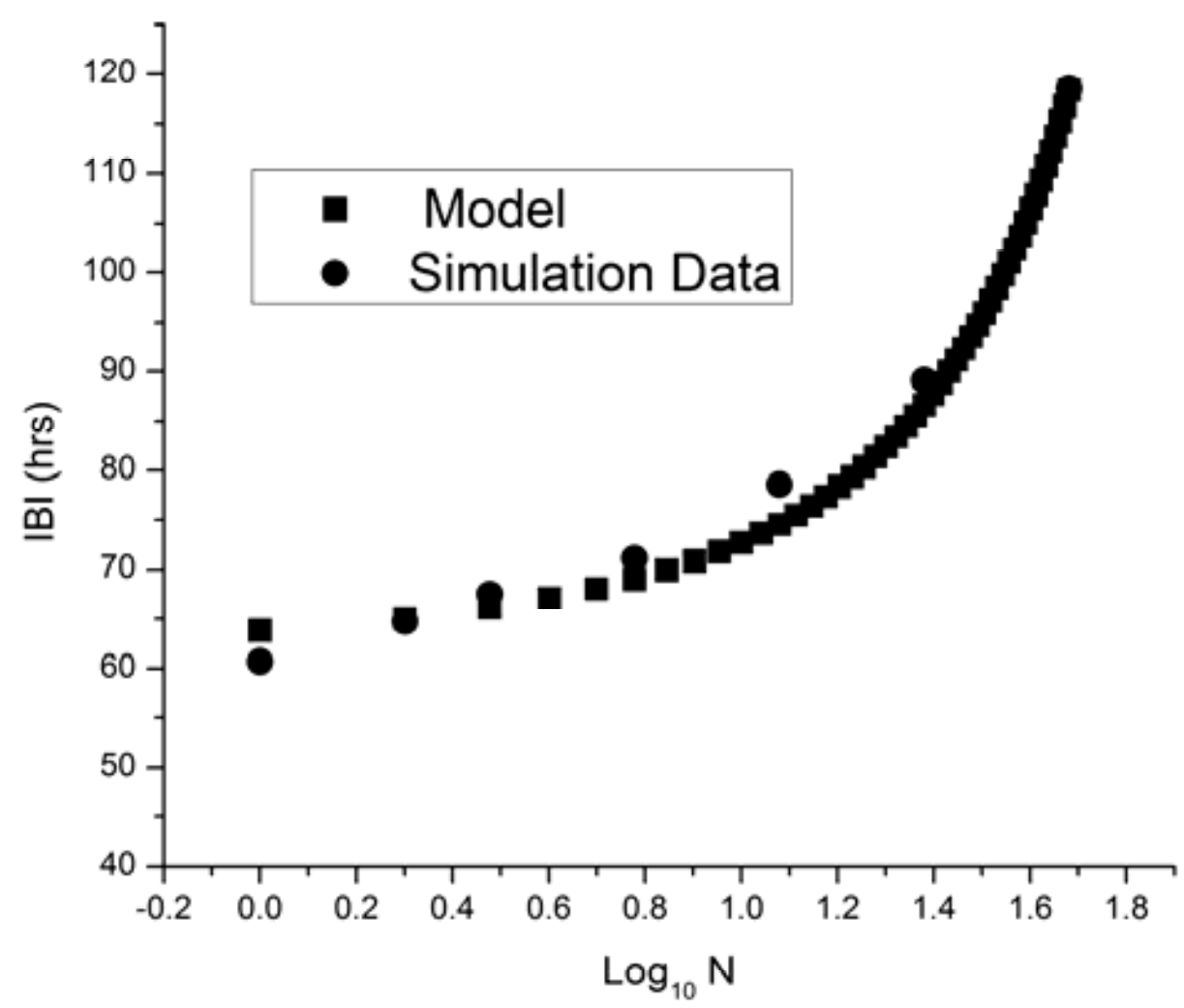

**Figure 10. The protein interburst interval (IBI) plotted as a function of the logarithm of the gene copy number.** Squares: results from Eq. (33) with $T_0$ = 56 hrs, $a$ = 0.35 and $\tau_0 = 15$ hrs. Circles: results from the stochastic simulation of the reactions in Eqs. (1)-(16) with the R protein degradation rate $\delta_R = 0.06$.

## Discussion

The results presented in this paper indicate that stochastic effects due to small gene copy numbers play an important role in the dynamics of oscillatory gene networks. Many of these networks are described by models whose behavior is "excitable." That is, in the absence of stochastic fluctuations, the model may predict time-independent steady state concentrations of mRNA and the associated proteins. However, when stochastic effects, primarily due to fluctuations associated with small gene copy numbers, are taken into account, the model may predict (more or less regular) bursts of mRNA and protein production. If the parameters of the model are close to the boundary between steady state behavior and oscillatory behavior for the deterministic (non-stochastic) model, the stochastic model may exhibit this burst behavior over a wide range of gene copy numbers. For large gene copy numbers, one would expect the behavior of the stochastic model to approach that of the deterministic model, but close to the boundary between steady-state and oscillatory behavior (a bifurcation), that convergence can be very slow.

From the biological perspective, one might argue that Nature allows the system parameters to evolve to be close to the bifurcation boundary because in that parameter region, the average time between mRNA and protein bursts and the regularity of those bursts is relatively independent of gene copy number and transcription and translation rates. Whether this conclusion is true in a system exhibiting a supercritical Hopf bifurcation will be explored in future work.

## Author Contributions

Conceived and designed the study: RH. Wrote computer code and performed the calculations: RH BB JM AP AS. Analyzed the data: RH BB JM AP AS. Wrote the paper RH BB JM AP AS.

# Appendix A

| **Parameter** | **Numerical Value** |
|---|---|
| $\theta_A$ | 50 h$^{-1}$ |
| $\gamma_A$ | 1 molecules $^{-1}$ h$^{-1}$ |
| $\theta_R$ | 100 h$^{-1}$ |
| $\gamma_R$ | 1 molecules $^{-1}$ h$^{-1}$ |
| $\alpha_A$ | 50 h$^{-1}$ |
| $\alpha'_A$ | 500 h$^{-1}$ |
| $\alpha_R$ | 0.01 h$^{-1}$ |
| $\alpha'_R$ | 50 h$^{-1}$ |
| $\delta_{M_A}$ | 10 h$^{-1}$ |
| $\delta_{M_R}$ | 0.5 h$^{-1}$ |
| $\delta_A$ | 1 h$^{-1}$ |
| $\beta_A$ | 50 h$^{-1}$ |
| $\beta_R$ | 5 h$^{-1}$ |
| $\gamma_C$ | 2 molecules $^{-1}$ h$^{-1}$ |

**Table 1. Numerical values of the parameters used in this study.** They are the same as those used by [26]. $\delta_R$, the R protein degradation rate, is used as a control parameter.

| *transMF* → gene copy number↓ | 0.5 | 1 | 2 | 3 |
|---|---|---|---|---|
| 1 | | 7.39±0.12 | 7.65±0.11 | 7.84±0.13 |
| 2 | 7.63±0.13 | 8.32±0..13 | | |
| 3 | | 8.02±0.17 | | |

**Table 2. Regularity of the protein burst intervals for several combinations of the product gene copy number × *transMF*.** The R protein degradation rate $\delta_R = 0.06$. Compare numbers with products of (gene copy number) × (*transMF*). Averages from 40 noise realizations.

## References

1. Kærn M, Elston TC, Blake WJ, Collins JJ. (2005) Stochasticity in gene expression: From theories to phenotypes. Nat. Rev. Genet. 6: 451-464.

2. Raser JM, O'Shea EK. (2005) Noise in gene expression: Origins, consequences, and control. Science 309: 2010-2013.

3. Süel GM, Kulkarni RP, Dworkin J, García-Ojalvo J, Elowitz MB. (2007) Tunability and noise dependence in differentiation dynamics. Science 315: 1716-1719.

4. Koseska A, Volkov E, Zaikin A, Kurths J. (2007) Quantized cycling time in artificial gene networks induced by noise and intercell communication. Phys. Rev. E 76(2): 020901(R)-1-4.

5. Turcotte M, García-Ojalvo J, Süel GM. (2008) A genetic timer through noise-induced stabilization of an unstable state. Proc Natl Acad Sci U S A 105(41): 15732-15737.

6. Ullner E, Buceta J, Díez-Noguera A, García-Ojalvo J. (2009) Noise-induced coherence in multicellular circadian clocks. Biophys J 96(9): 3573-3581.

7. Çağatay T, Turcotte M, Elowitz MB, García-Ojalvo J, Süel GM. (2009) Architecture-dependent noise discriminates functionally analogous differentiation circuits. Cell 139(3): 512-522.

8. Margueron R, Reinberg D. (2010) Chromatin structure and the inheritance of epigenetic information. Nat Rev Genet 11(4): 285-296.

9. Mazzio EA, Soliman KFA. (2012) Basic concepts of epigenetics: Impact of environmental signals on gene expression. Epigenetics 7(2): 119-130.

10. Gillespie DT. (1977) Exact stochastic simulation of coupled chemical reactions. J. Phys. Chem. 81(25): 2340-2361.

11. Lande R, Engen S, Sæther B. (2003) Stochastic population dynamics in ecology and conservation. Oxford; New York: Oxford University Press.

12. Kim HD, Shay T, O'Shea EK, Regev A. (2009) Transcriptional regulatory circuits: Predicting numbers from alphabets. Science 325(5939): 429-432.

13. van Kampen NG. (1992) Stochastic processes in physics and chemistry. Amsterdam: Elsevier.

14. Gillespie DT. (2000) The chemical langevin equation. J. Chem. Phys. 113(1): 297-306.

15. Murphy KF, Balázsi G, Collins JJ. (2007) Combinatorial promoter design for engineering noisy gene expression. Proc. Nat. Acad. Sci. 104(31): 12726-31.

16. Lu TK, Khalil AS, Collins JJ. (2009) Next-generation synthetic gene networks. Nat Biotechnol 27(12): 1139-1150.

17. Acar M, Pando BF, Arnold FH, Elowitz MB, Oudenaarden Av. (2010) A general mechanism for network-dosage compensation in gene circuits. Science 329(5999): 1656-1660.

18. Sudmant PH, Kitzman JO, Antonacci F, Alkan C, Malig M, et al. (2010) Diversity of human copy number variation and multicopy genes. Science 330(6004): 641-646.

19. Mills RE, Walter K, Stewart C, Handsaker RE, Chen K, et al. (2011) Mapping copy number variation by population-scale genome sequencing. Nature 470(7332): 59-65.

20. Torres EM, Williams BR, Amon A. (2008) Aneuploidy: Cells losing their balance. Genetics 179(2): 737-746.

21. Pavelka N, Rancati G, Zhu J, Bradford WD, Saraf A, et al. (2010) Aneuploidy confers quantitative proteome changes and phenotypic variation in budding yeast. Nature 468(7321): 321-325.

22. Springer M, Weissman JS, Kirschner MW. (2010) A general lack of compensation for gene dosage in yeast. Mol Syst Biol 6: 368-368.

23. Lee JA, Lupski JR. (2006) Genomic rearrangements and gene copy-number alterations as a cause of nervous system disorders. Neuron 52(1): 103-121.

24. Zakharova A, Kurths J, Vadivasova T, Koseska A. (2011) Analysing dynamical behavior of cellular networks via stochastic bifurcations. PLoS ONE 6(5): e19696-1-12.

25. Arnold L. (2003) Random dynamical systems. Berlin: Springer.

26. Vilar JMG, Kueh HY, Barkai N, Leibler S. (2002) Mechanisms of noise-resistance in genetic oscillators. Proc. Nat. Acad. Sci. 99(9): 5988-5992.

27. Gillespie DT. (1976) A general method for numerically simulating the stochastic time evolution of coupled chemical reactions. J. Comput. Phys. 22: 403-434.

28. Busino L, Bassermann F, Maiolica A, Lee C, Nolan PM, et al. (2007) $SCF^{Fbx13}$ controls the oscillation of the circadian clock by directing the degradation of cryptochrome proteins. Science 316: 900-904.

29. Galla T. (2009) Intrinsic fluctuations in stochastic delay systems: Theoretical description and application to a simple model of gene regulation. Phys Rev E 80(2): 021909-1-9.

30. Boland RP, Galla T, McKane AJ. (2009) Limit cycles, complex floquet multipliers, and intrinsic noise. Phys Rev E 79(5): 051131-1.

31. Steuer R, Zhou C, Kurths J. (2003) Constructive effects of fluctuations in genetic and biochemical regulatory systems. BioSystems 72(3): 241-251.

32. Hou Z, Xin H. (2003) Internal noise stochastic resonance in a circadian clock system. J. Chem. Phys. 119(22): 11508-11512.

33. Yeomans JM. (1992) Statistical mechanics of phase transitions. Oxford [England], New York: Oxford University Press.

34. Sethna JP. (2006) Statistical mechanics, entropy, order parameters, and complexity. Oxford, New York: Oxford University Press.

35. Hilborn RC, Erwin RJ. (2005) Fokker-Planck analysis of stochastic coherence in models of an excitable neuron with noise in both fast and slow dynamics. Phys. Rev. E 72: 031112-1-14.